\documentclass{skaox2006}
\usepackage{graphicx}
\newcommand{\ltsimeq}{\raisebox{-0.6ex}{$\,\stackrel
        {\raisebox{-.2ex}{$\textstyle <$}}{\sim}\,$}}
\newcommand{\gtsimeq}{\raisebox{-0.6ex}{$\,\stackrel
        {\raisebox{-.2ex}{$\textstyle >$}}{\sim}\,$}}
\begin{document}
   \title{Cosmology with the Square Kilometre Array}

   \author{Steve Rawlings\inst{1,2}
          }

   \institute{Astrophysics, University of Oxford, Denys Wilkinson Building, Keble Road, Oxford, OX1 3RH, UK \and sr@astro.ox.ac.uk}

   \abstract{ 
We review how the Square Kilometre Array (SKA) will address fundamental questions in cosmology, focussing on its use for neutral Hydrogen (HI) surveys. A key enabler of its unique capabilities will be large (but smart) receptors in the form of aperture arrays. We outline the likely contributions of Phase-1 of the SKA (SKA$_{1}$), Phase-2 SKA (SKA$_{2}$) and pathfinding activities (SKA$_{0}$). We emphasise the important role of cross-correlation between SKA HI results and those at other wavebands such as: surveys for objects in the EoR with VISTA and the SKA itself; and huge optical and near-infrared redshift surveys, such as those with HETDEX and Euclid. We note that the SKA will contribute in other ways to 
cosmology, e.g. through gravitational lensing and $H_{0}$ studies.  
}
   \maketitle
%
%
\section{Introduction}
\label{sec:intro}

Over the last two decades, observations of the Cosmic Microwave Background (CMB) have revolutionised our understanding of the Universe. They have
confirmed theoretical predictions that the seeds of all structure are present $\sim 0.4$ Myr ($z \sim 1100$) after the Big Bang, in the form of tiny ($\sim 1$ part in $10^{5}$) fluctuations in an otherwise smooth, featureless Universe. At this stage, gravity has had only a limited time to amplify these fluctuations, so their power spectrum $P(k)$, imprinted as an angular pattern in the CMB, has proven rich in cosmological information. This
includes an oscillatory (with scale) signature due to `Baryon Acoustic Oscillations'\footnotemark ,
\footnotetext{
BAOs are frozen-in plasma oscillations, or sound waves, from the pre-recombination Universe that, observed through cosmic time,
act as a cosmic `standard ruler'.}
and other features like the ratio of power on large and small physical scales that informs on neutrino mass\footnotemark .
\footnotetext{
Higher neutrino mass, and hence higher energy density in the known cosmic number density of neutrinos, means less clustering, and power on small scales, because neutrinos, a form of hot dark matter, `free-stream' out of Cold-Dark-Matter (CDM) condensations.
} 

Here we argue that the Square Kilometre Array (SKA, Rawlings \& Schilizzi \cite{Rawlings2011}) will be a key facility in answering the questions of
how the Universe formed its structure - galaxies, black holes, and stars. It will allow astronomers access to epochs between $z \sim 20-30$ and $z \sim 6$ when energy sources from the forming structure drove fundamental changes in the neutral Hydrogen (HI), ending in an Epoch of Reionization (EoR) around $z  \sim 10$ when the HI between galaxies became reionized (e.g.\ Zaroubi \cite{Zaroubi2011}). 

We will also
argue that the SKA  is poised to become the premier tool for probing the large-scale structure of the Universe after the EoR, using features in $P(k)$ and other statistical measures, to understand the properties of dark energy or post-Einstein gravity, to measure neutrino masses, and to study the causes of cosmic inflation. 

\section{Surveying the Universe}

Table~\ref{tab:one} gives the number of independent (Fourier) modes of the Universal density field around a comoving
wavenumber $k= 2 \pi / x = 0.125 ~ \rm Mpc^{-1}$ corresponding to a length scale $x=50 ~ \rm Mpc$. At $z=0$, this scale lies at the edge of, the `linear regime'  where the matter over-density $\delta \ltsimeq 1$, but, at higher $z$, it lies comfortably within the linear regime and is smaller than the first two `wiggles' in the critical BAO signature. Table~\ref{tab:one} considers various sky areas and redshift depths relevant to existing, or upcoming, cosmological surveys; similar types of data are presented graphically in Loeb \& Wyithe (\cite{Loeb2008}).

\begin{table*}
$$
 \begin{array}{cclll}
            \hline
            \rm{Range~in~}z  & \Omega (\rm{sr}) & N_{\rm modes} & \delta P / P & \rm{Surveys} \\
            \hline
            0.0 - 0.2 & 3.0 & 3 \times 10^{4} & 6 \times 10^{-3} & \rm {SDSS, SKA}_{0} \\ 
            0.2 - 0.7 & 3.0 &  8 \times 10^{5} &  1 \times 10^{-3} & \rm{BOSS} \\
            0.2 - 2.0 & 0.06 & 1 \times 10^{5} &  3 \times 10^{-3} & \rm{SKA}_{1} \\
            0.2 - 2.0 & 6.0 &  1 \times 10^{7}  &  {\bf 3 \times 10^{-4}}  & \rm{SKA}_{2}, \rm{BigBOSS}, \rm{Euclid} \\
            2.0 - 3.0 & 0.3  &  6 \times 10^{5}   & 1 \times 10^{-3}  & \rm{HETDEX} \\
            2.0 - 6.0 & 0.01 &  7 \times 10^{4}  &         ^a                     & \rm{SKA}_{1} \\ 
            2.0 - 6.0 & 6.0  &  4 \times 10^{7}  &    \bf{2 \times 10^{-4}}  & {\bf \rm{SKA}_{2}} \\
            6.0 - 13.0 & 0.03 &  2  \times 10^{5} &      ^b                        & \rm{SKA}_{0} \\
            6.0 - 13.0 & 0.03 & 2 \times 10^{5} &  2 \times 10^{-3}  & \rm{SKA}_{1} \\
            6.0 - 13.0 & 3.0 &  2  \times 10^{7} &  \bf{2 \times 10^{-4}} & {\bf \rm{SKA}_{2}} \\
             13.0 - 30.0 & 0.03 & 2  \times 10^{5} &    ^b                          & \rm{SKA}_{1} \\
             13.0 - 30.0 & 3.0 & 2  \times 10^{7} & \bf{2 \times 10^{-4}} & {\bf \rm{SKA}_{2}} \\
             \rm{CMB} & 11.0^c &  2  \times 10^{5} &                               & \rm{WMAP, Planck} \\ 
            \hline         
\end{array}
$$
{\caption[junk]{\label{tab:one}
{
Measures of the density field in the Universe possible with various upcoming surveys (those with power-spectrum accuracies marked in bold will allow transformational progress). 
Column 1: redshift range of spectroscopic survey. 
 Column 2: sky area of survey (in sr). 
 Column 3:  number of independent Fourier modes $N_{\rm modes}$ in the survey (covering cosmic comoving volume $V$) over a factor of e spread in $k$ centred on $k=0.125 ~ \rm Mpc^{-1} $ (equivalent to 50 Mpc in comoving units); this is calculated using  $N_{\rm modes} = [V / (2 \pi)^3] \times 2 \pi k^3$.  
 Column 4: Fractional error in measured power spectrum $P(k)=< |\delta_{k}|^2 > $ (where $\delta_{k}$ are the Fourier coefficients of the transform of the real-space over-density $\delta = \delta \rho / \rho$), given by $1 / \sqrt{N_{\rm modes}}$, assuming the survey is limited by cosmic variance (rather than shot or experimental noise) - this column is left blank whenever the example surveys are not in this regime (or are essentially 2D surveys, as is the case for the CMB); in practice, $P(k)$ will be measured in several, say 4, redshift bins, with correspondingly lower accuracy, say factor $\sim$ 2, in each bin.    
Column 5: survey mentioned in the text. Although there are other planned ground-based optical surveys, we consider HETDEX, 
BOSS
and BigBOSS as examples;  for space-based near-IR redshift survey data, we consider
Euclid.
}}}
        \begin{list}{}{}
\item[$^{\mathrm{a}}$] Cosmic-variance-limited mapping will become possible over a limited redshift range provided the SKA Advanced Instrumentation Program (see Rawlings \& Schilizzi \cite{Rawlings2011}) generates the capability for early deployment of a small fraction of the SKA$_2$ mid-frequency AAs in a new core embedded within SKA$_{1}$ (see Section~\ref{sec:phased}).
\item[$^{\mathrm{b}}$] A large amount of mode averaging is needed to obtain a statistical detection of the structure encoded in $P(k)$. 
\item[$^{\mathrm{c}}$] Assuming $\sim 10$ per cent of the sky is unusable 
because of residual Galactic Plane contamination.
\end{list}
   \end{table*}
   
The surveys given as examples in Table~\ref{tab:one} probe the density field in different ways. 
Optical and near-IR redshift surveys (SDSS, BOSS\footnotemark 
\footnotetext{http://www.sdss3.org/surveys/boss.php/}, BigBOSS\footnotemark 
\footnotetext{http:/bigboss.lbl.gov/}, Euclid [Laureijs et al.\ \cite{Laureijs2008}], 
HETDEX [Hill et al.\ \cite{Hill2008}]), and `thresholded'\footnotemark
\footnotetext{
Catalogued surveys formed from detecting `islands' of HI emission at $n \sigma$, with $n \sim 5-10$, above an experimental background with r.m.s. $\sigma$.
}
SKA surveys detect individual galaxies that are taken to Poisson-sample the underlying field. For these surveys, regions where
$n V P(k) \gtsimeq 1$ ($n$ is the comoving density of survey galaxies) is the condition for shot noise to be sub-dominant, and regions where $n V P(k) \gg 1$, allow the capability of checking for systematic errors by  comparing results from different sub-sets of the data. 

\begin{figure*}
\vspace{250pt}
\includegraphics{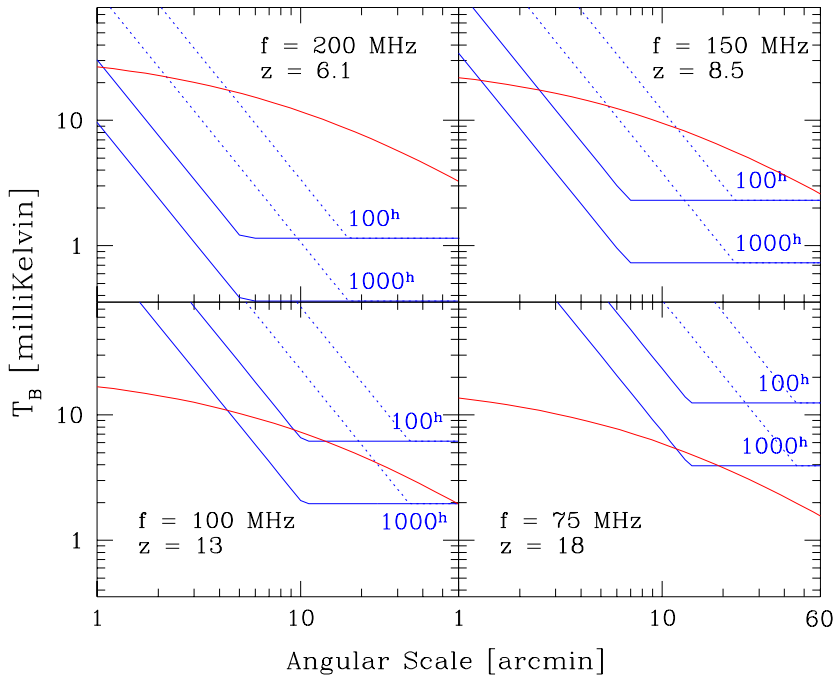}
\caption{
Estimated temperature of HI fluctuations as a function of angular scale 
reproduced from Furlanetto \& Briggs (\cite{Furlanetto2004}). The gently curved red lines represent 3$\sigma$ intrinsic HI fluctuations in $\delta \nu = 0.5$ MHz bins assuming the IGM is neutral, e.g. $x_{H}=1$ [clearly a bad assumption at $z=6.1$ where a value $x_{HI} \sim 0.02$ would scale the red line down by a factor $\sim 50$]. The solid lines show
1$\sigma$ sensitivities predicted in 100- and 1000-hour exposures that lie very close to predictions for SKA$_{1}$ (the dotted lines should be ignored). Real data from the GMRT (Paciga et al. \cite{Paciga2010}) have produced upper limits at the $\sim 100$ mK level. 
}
\label{fig:AA}
\end{figure*}

CMB and `non-thresholded' SKA surveys integrate all contributions to temperature (surface brightness) fluctuations $\delta T / T$ in the sky\footnotemark . In the case of the SKA surveys this works by adding up all the contributions from HI emission inside and between galaxies, regardless of any thresholding criteria: after reionization, this probably still amounts to counting galaxies, as neutral HI can only exist in the dense self-shielding environments in or near galaxies.
\footnotetext{
The contributions to $\delta T / T$ in the CMB and (SKA) HI signals are complicated: in the case of CMB because there are so many
primary and secondary cause of CMB fluctuations; in the case of HI because there are many astrophysical processes capable of generating HI
anisotropies. Both CMB and (SKA) HI  experiments will have to deal with the subtraction of bright, complex and polarized foreground emissions: the successes of CMB astronomy in overcoming such difficulties provide both hope and lessons for HI studies with the SKA.
}

Thus, analysis of non-thresholded SKA data, or `HI intensity mapping'  (Chang et al.\  \cite{Chang2008}; Chang et al.\ \cite{Chang2010}), 
has the potential to utilize many more modes of the density field than a thresholded survey: much larger signals in the  $n V P(k) <  1$ regime becomes available in the form of summed HI emission modulated by the cosmic web of large-scale structure. The fluctuations around the absolute brightness temperature level $T_{b}$ of the HI sky is given by (Furlanetto et al.\ \cite{Furlanetto2007})

\begin{equation}
\label{eqn:one}
\delta T_{\rm b} \sim
9 \delta  x_{H} 
\frac{T_{\rm S} - T_{\rm CMB} }{T_{S} } 
(1+z)^{1/2}
\frac{H(z) / (1+z) }{ \partial V_{r} / \partial r } ~ \rm mK,
\end{equation}

\noindent
where $x_{H}$ is the fraction of Hydrogen that is neutral, and the fraction involving spin temperature $T_{\rm S}$ and CMB temperature
$T_{\rm CMB}$ saturates (at unit value) once the IGM has been heated by astrophysical sources; the last fraction, involving the Hubble parameter $H(z)$ and the radial gradient in radial velocity, accounts for the more subtle effects of redshift-space distortions. After reionization, the power spectrum of fluctuations traces the power spectrum of the underlying fractional over-density $\delta$ in the matter power spectrum. During reionization, on small (arcmin, or comoving-Mpc) scales, the signal level is boosted by the formation and percolation of bubbles of ionized material around photon-producing sources and other 
effects (Furlanetto et al.\  \cite{Furlanetto2007}). After reionization, measurements of the HI associated with the damped Ly$\alpha$ absorption lines of quasars
suggests $x_H \sim 0.02$ with no strong dependence on redshift (Trenti \& Stiavelli \cite{Trenti2006}).

Roughly speaking, once the EoR ends at $z \ltsimeq 6$, the neutral fraction $x_{H} \sim 0.02$, so Equation~\ref{eqn:one} shows that the modulation (e.g.\ of density fluctuations of amplitude $\delta \sim 1$ around just-collapsing structures) produces temperature fluctuations with an amplitude $\sim 500 [(1+z) / 7]^{1/2}~ \rm \mu K$. This is in good agreement with values inferred from the cross-correlation of HI and optical data by Chang et al.\ (\cite{Chang2010}) at redshift $z \sim 0.8$. During reionization,  detailed models (e.g. Furlanetto et al.\ \cite{Furlanetto2007}), constrained by CMB (WMAP) measurements of the column of
ionized material towards the CMB, suggest amplitudes $\sim 10 ~ \rm mK$. 

EoR experiments are very challenging as they must understand
their error budgets sufficiently to assign any `excess' (above thermal noise) variance in their data to intrinsic 
HI fluctuations rather than, e.g., RFI or residuals from ionospheric 
calibration and foreground removal. As has proven to be the case in attempts to make statistical detections of HI at more modest redshifts, e.g. $z \sim 0.8$ (Chang et al.\  \cite{Chang2010}), convincing `auto-correlation detections'\footnotemark ~of HI 
\footnotetext{
In an (RA, DEC, $z$) data cube from any radio telescope with well understood sources of (non-astrophysical) variance, the normal ergodic assumption means that variance calculated statistically across spaxels is equivalent to an `auto-correlation' $< \rho \times \rho>$ where $\rho$ represents the signal in each spaxel. A cross-correlation signal, e.g. $< \rho_{1} \times \rho_{2}>$ from two independent data cubes, e.g. a radio HI cube with a signal 
(in each spaxel) $\rho_{1}$ and an optical galaxy density cube with signal (in each spaxel) $\rho_{2}$, is much less prone to systematics as sources of variance (other than astrophysical signal) should be independent and average to zero in the cross-correlation. Note also that higher-order statistics, such as skewness, may helpful in extracting true HI signal (Harker et al.\ 2009). 
}
variance may be much harder 
to obtain than `cross-correlation' detections where the uncorrelated errors in two ways of probing the EoR may be exploited to obtain robust
measurements of the astrophysical signals\footnotemark . 

It should also be noted that HI power-spectrum methods are not yet proven on astrophysical datasets, and being inherently statistical in nature, will be challenging to establish. Thresholded surveys allow the use of the data itself to check for systematics or, through measured redshift space distortions, to cleanly marginalise over galaxy bias (the scale factor between fluctuations in HI and fluctuations in dark matter, see also Abdalla et al. \cite{Abdalla2010}).

\footnotetext{
There are astrophysical signals beyond the Milky Way that must be removed (e.g. Jelic et al.\ \cite{Jelic2008}), and cross-correlation methods may be useful in separating these from the HI signal: the radio background due to 'foreground' radio sources is more than three-orders-of-magnitude brighter than the 
HI signal (Bridle \cite{Bridle1967}), and not smooth on the sky or in redshift due to clustering; this source of variance will be strongly correlated with the location of galaxies in deep near-infrared images.
}
\section{The phased construction of the SKA and its emerging role in cosmology}
\label{sec:phased}

The affordability of building and operating the SKA is strongly dependent on the solutions adopted for the signal and 
data transfer and processing, and strongly justifies the decision to build the SKA in phases. This argument is particularly clear for HI
cosmology experiments that involve both the aperture array (AA) and dish parts of the SKA. Other arguments regarding the phased construction of SKA dishes come from the requirements of pulsar astronomy (Kramer \cite{Kramer2011}). 
For convenience we use the term SKA$_{0}$  here to refer to scientific and technical verification 
results imminent from the SKA pathfinder programs [see Rawlings et al.\ (\cite{Rawlings2011}) for a discussion of some of the key pathfinder programs, and for descriptions of the SKA$_{1}$ and SKA$_{2}$ realizations assumed herein].

The necessary evolution in AAs from SKA$_{0}$ to SKA$_{1}$ is best appreciated by studying an 
aerial photograph of the LOFAR core\footnotemark . This shows a `superterp' of diameter $D_{\rm core} = 0.34$ km, but, at 130 MHz, it is obvious from the small density of high-band antennae stations that the array has a low filling factor ($\mu_{\rm f} \sim 0.02$ out to $r=0.5$ km) showing that pathfinder projects are approaching, but still some way off, the scale of the SKA$_{1}$ AA core. Critically, however, LOFAR is already being used for both astronomy and technology verification. The SKA$_{1}$ core will improve on SKA$_{0}$ in a number of ways: (i) it will be situated in a radio-quiet zone in Australia or South Africa, minimising the need for RFI flagging of the data [which can greatly complicate statistical measurement of the HI power spectrum as it is a source of data variance along the frequency, or redshift, axis] and minimising the number of bits required in the early digitisation [a construction and operating cost driver for the SKA] (ii) It will, with $D_{\rm core}=$1 km, have $\sim 10$-times the physical area of the superterp; (iii) it will have a much higher filling factor ($\mu_{\rm f} \sim 0.8$ c.f. $\mu_{\rm f} \sim 0.02$); and (iv) its methods of beam forming and correlating signals may be entirely in the digital domain, allowing greater control of systematic errors and calibration. These points mean that in terms of temperature sensitivity at 130 MHz (HI at $z=10$) and 9-arcmin (25 Mpc comoving) resolution, SKA$_{1}$ will outperform SKA$_{0}$ by a factor $\sim 100$, making it the first instrument capable of high-fidelity imaging of both the intrinsic EoR fluctuations and the foregrounds that must be accurately removed to reveal them.

\footnotetext{
http://www.lofar.org/about-lofar/image-gallery/latest-lofar-images/
}

\subsection{The Dark Ages and the EoR}

The information in the radio sky arrives on Earth as waves of wavelength $\lambda$ from all directions above the local horizon. With a future detection of HI signal from the EoR as the astronomers' first step into the dark ages, SKA$_{0}$ interferometer experiments are already interrogating this information to seek statistical detections of HI fluctuations in the redshift range $6 \ltsimeq z \ltsimeq 13$\footnotemark
\footnotetext{
Other experiments, e.g. Bowman \& Rogers (\cite{Bowman2010}), are using much smaller numbers of receivers targeted at absolute, rather than differential, detection of HI, in the form of a monopole step-change in sky brightness with redshift due to reionization. They also target the $z=13-6$ range as this comfortably brackets uncertainties from CMB studies regarding when the Universe transitioned from $x_{H} \sim 1$ to $x_{H} \sim 0.02$.} 
Fig.\ \ref{fig:AA} illustrates that the `sweet-spot' for detection and imaging the EoR must be somewhere near the centre of this redshift range: at $z \sim 6$ the IGM is known (from optical absorption features in quasars) to be almost entirely ionized (Ouchi \cite{Ouchi2011}), and by $z \sim 13$ (100 MHz) the sky temperature rises are starting to strongly degrade the signal-to-noise ratio of any proposed experiment. Also clear from Fig.\ \ref{fig:AA}  is that at an angular scale around $\sim 9$ arcmin (25 Mpc, comoving, at $z \sim 10$) is optimum as, for a fixed number of receiver chains, there is a rapid loss of temperature sensitivity towards smaller angular scales, tensioned against a gentler drop in HI signal level towards larger angular scales. 

To image the sky to detect HI at $z \sim 10$ (HI redshifted to $\lambda = 2.3$ m or 130 MHz)
with high fidelity, it is necessary to space (many of) the AA elements by $\sim \lambda / 2 = 1.15$ m so as to 
form a dense aperture array that Nyquist samples the incident wavefront and hence avoids `grating lobes' above the horizon. The central ($D_{\rm core}=1$-km diameter) core of the AA part of the SKA can then be considered as a single large (smart) aperture that has a natural FOV (or `core beam') of angular width $\approx 1.22 \lambda / D \sim  9$ arcmin that, by suitable addition of phase-weighted
signals from the antennae, can be one of  many beams [these can increase FOV and, potentially, adaptively null sources of RFI], or by suitable cross-correlation of antenna signals can continuously map out bright sources, and hence monitor ionospheric conditions across the sky.

The SKA$_{0}$ AA experiments are competing to obtain the first statistical detection of the HI EoR fluctuations. These all aim to 
achieve this by averaging together all the independent Fourier modes of the HI distribution that they can measure in the plane of the sky and in the frequency, or redshift, direction, to measure the power-spectrum (scale dependence) of the HI fluctuations. In principle, the gains of a power spectrum approach are huge (see Table~\ref{tab:one}), with, say, considering 25-Mpc (9 arcmin) modes over a survey of 20 $\rm deg^{2}$ (which is comfortably within the FOV of the 
LOFAR and MWA high-band analogue beam formers) giving, transverse to the line-of-sight, $\sqrt{N_{\rm modes}} \sim \sqrt{900}=30$ that, averaging also over a cube (in comoving coordinates) in the redshift direction give a total mode-averaging boost in total signal-to-noise ratio of $\sim 165$. This gain can grow further with additional redshift binning or
sky coverage, explaining why statistical detections remain highly plausible 
(e.g. Zaroubi \cite{Zaroubi2011})
despite the low temperature sensitivities ($\mu_{\rm f} \gtsimeq 0.01$ over $D_{\rm core} = 1$ km) of SKA$_{0}$ experiments (Fig.\ref{fig:AA}).

There is also the possibility of exploiting cross-correlation techniques in the EoR. A stacking analysis\footnotemark
\footnotetext{A stacking analysis - cross-correlation of the radio data with a set of 3D Dirac Delta functions centred on the (RA, DEC, $z$) coordinates of 
objects from other surveys - is the simplest variant of such methods. These methods will be particularly challenging in the case of Ly$\alpha$ surveys because resonant HI absorption strongly affects the line profiles and centroids, and requires a sophisticated stacking technique.} 
could be based on Ly$\alpha$-emitting galaxies (e.g. Ouchi \cite{Ouchi2011}): an advantage here is that the advent of wide-field narrow-band near-IR capabilities with telescopes such as VISTA should deliver 1,000s of $z \sim 7$ objects across EoR fields over the coming years. The disadvantages of this approach include: a limited likely redshift span of large samples of objects due to the increased rarity and faintness of suitably bright Ly$\alpha$ emitters at higher $z$; and likely redshift-dependent switches in the sign of the expected cross-correlation signal. For example, at $z \sim 7$, the positive correlation expected because HI emission is concentrated in galaxy haloes, reverses to a negative correlation at $z \sim 10$ when ionized bubbles begin to percolate the Universe, and the HI signal may peak away from the sources of Ly$\alpha$ photons and heat. We emphasise that achieving an angular resolution of at least the $\sim 1-$arcsec level of seeing-limited optical data is critical because only then can sign-switches in cross-correlation analyses be used as useful information rather than as a source of confusing variance.

Another potential cross-correlation method in the EoR is between HI and gas-rich galaxies detected by radio telescopes at high frequency: e.g. through the 
molecular CO(1-0) line with SKA$_{0}$.
This is discussed by Heywood et al.\ (\cite{Heywood2011a}) but we emphasise here that the dish-part of SKA$_{1}$ may be needed to get 1000s of tracers across the EoR fields with a broad range of redshifts where galaxies are present and detectable; this requires SKA$_{1}$  dishes to have good efficiency up to at least 15 GHz [CO(1-0) at $z \sim 6.5$]. Again, the HI data cubes will need $\sim$1-arcsec 
resolution for efficient cross-correlation experiments. There are plans to construct new instruments aimed at adding `CO intensity mapping'
(e.g. Gong et al.\ \cite{Gong2011}; Lidz et al.\  \cite{Lidz2011}) to the armoury of techniques for studying the EoR. We also note plans to cross-correlate HI signals with CMB data (e.g. Tashiro et al.\ \cite{Tashiro2010}) that would reveal effects such as 
an expected correlation between regions of low HI column density, and hence high ionized density, with large CMB optical depth.

The various SKA${_0}$ low-frequency-AA experiments are also comparing and contrasting various technical solutions relevant to the working of a `large-but-smart' highly-filled aperture, but there are important implementation details rather than 
technical show-stoppers holding up plans to build an SKA$_{1}$ core. SKA$_{1}$ will move the problem into a new regime where the HI signal targeted has fluctuations that on the relevant angular scales exceed the fluctuations due to thermal noise. The scientific and technical thinking behind the necessary leap in sensitivity required to go from cross-correlation, or tentative auto-correlation,
detections of the EoR (all that is possible with SKA$_{0}$) to an imaging instrument are illustrated in Fig.\  \ref{fig:AA}. 
The key point is that, in feasibly long exposures with SKA$_{1}$, astronomers will reach the regime where the signal-to-noise ratio of HI fluctuations and the
foregrounds are all much greater than unity, meaning that direct mapping of EoR HI features 
becomes much more tractable in the presence of polarized foregrounds, ionospheric effects and RFI. If SKA$_{0}$ does not yield
any definitive auto-correlation detections of HI in the EoR, constructing SKA$_{1}$ will be the only way forward in this research area; 
if SKA$_{0}$ does detect HI EoR statistically, via auto- or cross-correlation techniques, the lessons of CMB astronomy tell us that instruments capable of 
directly imaging the HI fluctuations will be urgently needed and rapidly exploited.

The scaling up from the AAs in SKA$_{0}$ crudely follows `Moore's law' in that from 2008 (the start of the construction of the LOFAR core) 
to the start of construction of the SKA$_{1}$ core in 2016 - i.e. roughly 7.5 years, or 5 lots of 18 months - the required 
number of antennae should be able to grow by a factor $\sim 2^{5} = 32$ that comfortably exceeds the increase in core antennae: from $\sim 18,400$ LOFAR high-band dipoles to $\sim 280,000$ SKA$_{1}$ dipoles. There will be increased central concentration of the collecting area, bringing with it other major cost savings on data transport and infrastructure that should allow an all-digital solution (retiring the need for analogue beam-forming that feeds the outputs of 16 antennae in LOFAR and MWA high-band tiles to a single receiver chain). The SKA project has developed sophisticated cost-estimation tools (Bolton et al.\ \cite{Bolton2009}) that lends great confidence that such an AA system can be built and operated within the SKA${_1}$ budget restrictions.

Fig.~\ref{fig:AA} shows that SKA$_{1}$ will have sufficient temperature sensitivity in a 1000-hr exposure to image fluctuations. The most efficient and reliable
way of covering each of a few, say 5, $20 ~ \rm deg^{2}$ patches is to build a 
hierarchical, but preferably still all-digital, beam-forming and correlation solution: following Dewdney et al. (\cite{Dewdney2010}), with 180-m diameter stations, each with a 50-arcmin station beam, requires only $N_{\rm station} \gtsimeq 30$ to ensure continuous sky coverage with station-beam overlap, and outrigger beams for calibration and RFI excision. With 25 of 50 AA stations in the core (radius $r < 0.5$ km), the core filling factor of $\mu_{\rm f} = 0.8$ (c.f. $\mu_{\rm f} \sim 0.01$ for SKA$_{0}$) provides excellent temperature sensitivity on large angular scales, while 10 further stations in the `inner' region  ($0.5 < r < 2.5$ km) will give sensitive imaging capability at resolutions down to $\sim 2$ arcmin (see Fig.\ \ref{fig:AA}); 15 further stations in 3 spiral arms out to $r=100$ km provides the resolution essential for the removal of point and extended foreground sources. Table~\ref{tab:one} shows that SKA$_{1}$ will produce a cosmic-variance-limited measurement of the power-spectrum at $z \approx 10$ that is comparable in accuracy to measurements of $P(k)$ from SDSS in the local Universe. SKA$_{1}$ will be able to map
out the evolution of the HI signal over the full range $6 \ltsimeq z \ltsimeq 13$: probing rapid changes in $x_{\rm H}$, as well as other factors in 
Equation~\ref{eqn:one},  with cosmic time will mean that astronomers will learn huge amounts about  the processes happening during this key epoch.
However, despite significantly enhancing knowledge of the reionization processes (and its effect on CMB data), this experiment is unlikely to produce game-changing measures of cosmological parameters. 

An EoR-optimised AA in SKA$_{1}$ will also be used to attempt statistical detections of HI at $z>13$, but the 
$T_{\rm sys} \sim T_{\rm sky} \lambda^{2.55}$ scaling at low frequency makes this a challenge. Putting aside the challenges of ionospheric calibration and foreground removal, that also worsen considerably with increasing $\lambda$, imaging looks challenging at $z >13$ (Fig.\ \ref{fig:AA}). However, simulations (Santos et al.\ \cite{Santos2011}) predict that the fluctuation signal may be boosted from the $\sim10$ mK EoR level (Fig.\ \ref{fig:AA})  to, perhaps, the $\sim 100$ mK level, meaning it might be possible to measure the HI power spectrum in the Dark Ages. Adopting the conservative prediction for the $z=18$ signal in Fig.\ ~\ref{fig:AA}, and realising that it would only take a few station beams to map, say, $20 ~ \rm deg^{2}$, then by mode averaging, it should be possible to attempt significant statistical detections across a few independent sky patches. These observations are critical for understanding how the HI signal reflects the first stars and black holes ahead of the EoR: measuring, and potentially even mapping, the predicted $\sim 100$ mK level fluctuations that would confirm the underlying conjecture of strong highly-spatially- clustered HI absorption due to cold ($T_{s} \ll T_{CMB}$; see Equation~\ref{eqn:one}) absorbing material near to the first galaxies (Barkana \& Loeb \cite{Barkana2004}),

In EoR and Dark Ages studies, the chief SKA$_{2}$ science drivers would then be to move from the limited resolutions, sky areas and redshift ranges observable with SKA$_{1}$: higher resolution is needed to map ionized structures directly associated with quasars and star-forming galaxies; covering a large fraction of the sky is essential for power-spectrum sensitivity and cross-correlation with CMB; and $13 \ltsimeq z \ltsimeq 20-30$ (100 MHz down to $\sim 70-45$ MHz) provides new information on the Universe. These can all be achieved simultaneously by, in SKA$_{2}$, building out the high-filling-factor AA from the core ($r < 0.5$ km) into the inner region ($0.5 < r < 2.5$ km), so that with long (1000 hr) exposures, there is sufficient sensitivity on few-arcmin scales to image $\sim 1$ mK fluctuations in  the EoR, and work, at least statistically, in the Dark Ages. 

This should delineate various stages expected in the Dark Ages: strong absorption associated
with the first stars; hot IGM, and hence HI emission, ($T_{S} \gg T_{\rm CMB}$, see Equation~\ref{eqn:one}) near the first black holes, and cold absorbing IGM surrounding these regions (Pritchard \& Furlanetto \cite{Pritchard2007}); and clear views of the subsequent process of reionization. For EoR fluctuations, SKA$_{2}$ will have a mapping speed gain of $\sim 25$ from sensitivity and $\sim 4-10$ from extra beams (and hence FOV)
sufficient to obtain cosmic-variance limited imaging over a significant fraction of the sky.
Table~\ref{tab:one} shows the low error bars on $P(k)$ that will result.
It is worth noting that because of the large sky temperatures at these low frequencies, that the SKA$_{2}$ core may need to have a collecting area approaching $\sim 10$ km$^{2}$! 

The removal of polarized Galactic foreground from EoR and Dark Age observations will need broad wavelength coverage to account for complex angular-scale dependent effects such as Faraday rotation and depolarization, and, together with the demands of pulsar surveys (Kramer \cite{Kramer2011}),  underpin the required SKA$_{1}$ ($\sim 70-450$ MHz) frequency range. It is therefore inevitable that SKA$_{1}$ will address many of the goals of the SKA magnetism key science project (Gaensler et al.\ \cite{Gaensler2004}).

\subsection{The post-EoR Universe}

In the post-EoR Universe, the dish-based arrays combined here under the term SKA$_{0}$ will make great strides towards understanding how the  the HI
in galaxies traces the underlying dark matter: e.g. ASKAP and WSRT/APERTIF are likely to generate all-sky HI surveys approaching in size the $\sim 
10^{6}$ galaxy surveys that current optical surveys like SDSS have delivered;  `deep' (to $z \approx 0.2$) 
surveys across regions heavily studied by optical redshift and near-IR imaging surveys, will allow sophisticated
cross-correlation analyses. Such efforts will be complemented by deep HI 
stacking and absorption-line experiments with MeerKAT, so the nature of HI evolution between $0 \ltsimeq z \ltsimeq 1$ may be moderately well understood ahead of SKA$_{1}$ operation. 

A first goal of the dish part of the SKA$_{1}$ will be a (largely thresholded) survey of HI galaxies between $z=0.2$ and $z \ltsimeq 2$, requiring 1000 hours 
of exposure. Assuming FOV extension from Phased-Array Feeds (PAFs) from the SKA Advanced Instrumentation Program (AIP) by a factor $\sim 10$ over that of a single-pixel feed at 700 MHz ($z=1$ HI), then 40 deg$^{2}$ of sky coverage (independent of frequency) per patch is plausible, yielding 200 deg$^{2}$ in five independent sky patches. 
Whilst not competitive with BOSS in terms of volume (see Table~\ref{tab:one}), the overlap regions between SKA$_{1}$ and BOSS 
would establish any limitations  imposed on either method due to systematic errors. 

Another main science driver for SKA$_{1}$ will be to push this type of detailed radio-optical cross-correlation work to the $z \sim 2-3$ regime. Here, the next-generation optical redshift surveys (e.g. HETDEX, Hill et al.\ \cite{Hill2008}) will be attempting to assemble the first large ($\sim 10^{6}$-galaxy-sized) samples. 
These are
critical epochs because they probe before dark energy is thought to make a major contribution to Universal expansion, so that  other critical cosmological parameters, such as the Universe's intrinsic curvature, can be constrained independently. In the equatorial overlap between (the mostly Northern Hemisphere) HETDEX and an AA survey with SKA$_{1}$\footnotemark , HETDEX  expects to discover $\gtsimeq 2 \times 10^{5}$ objects in the redshift range $2 \leq z \leq 3$ over $\gtsimeq 100 ~ \rm deg^{2}$ near the celestial equator. A 1000-hr SKA$_{1}$ exposure with SKA$_{1}$ achieves, at 450 MHz ($z = 2.1$) using AAs, $\gtsimeq 1 \sigma$ detections of  galaxies near the break of the HI mass function. A sky area of $\sim 100 ~ \rm deg^{2}$ could plausibly be covered instantaneously by AAs (if $N_{\rm beam} = 1500$).
\footnotetext{
HETDEX does not have continuous sky coverage, so a truly combined survey is plausible as the $\sim 15-$arcmin size of the HETDEX `tiles' is well matched to the low-frequency-station-beam size of SKA$_{1}$ at $\sim 450$ MHz.
A design consideration for all AAs in SKA is that they are able to observe the celestial equator due to the concentration of other wavelength 
surveys (such as HETDEX) and capabilities there.} With, conservatively, $10^5$ Ly$\alpha$ objects, and stacking techniques, the signal-to-noise ratio for HI  at $z \sim 2$ would be $\sim 300$, yielding the ability to split the sample by galaxy type (e.g.\ using near-IR estimates of stellar, and hence, dark matter mass). 

A direct measurement of the evolving HI galaxy power spectrum over at least some of the $z \sim 2-6$ range is also plausible and critical to pursue because Table~\ref{tab:one} shows that there are many modes available for HI intensity mapping with SKA$_{1}$. Over this range of redshifts, the temperature of the fluctuations on a fixed comoving scale are expected to be roughly constant. The `bias'\footnotemark
\footnotetext{
The bias $b$ is defined by $b^{2}$, with units of mK$^{2}$, as the ratio of the power spectrum of the temperature fluctuations to the product of the matter power spectrum and the square of the cosmic growth factor $g \approx 0.8 (1+z)$.
} is inferred from Equation~\ref{eqn:one} and should increase from $\sim 200 ~ \rm \mu K$ at $z \sim 2$ to $\sim 500 ~ \rm \mu K$ at $z \sim 6$. So, on a fixed comoving scale (say 50 Mpc), this compensates largely for the corresponding factor $\sim$3 decline in expected signal due to the growth [at $z  \sim 2$, the  temperature fluctuations on $\sim50$ Mpc scales will be $\sim 200 ~ \rm \mu K \times \sigma_{8} / g \sim 70 ~ \mu \rm K$ (taking $\sigma_{8} = 0.8$ to be the effective normalization of the power spectrum on the relevant scale)]. Note, again, the observational confirmation of this rough scaling at $z \sim 0.8$ 
(Chang et al.\  \cite{Chang2010}).

Assuming good progress from the SKA AIP, a relatively small (say 15 station) SKA$_{2}$-capable mid-frequency AA could be available near the time
of SKA$_{1}$ first operation With stations of diameter 56-m embedded in the SKA$_{1}$ configuration, and with antennae densely packed below $\sim 500-800$ MHz (Schilizzi et al.\  \cite{Schilizzi2007}) these would overlap in frequency with the SKA$_{1}$ low-frequency AAs over the redshift range $2.1 \ltsimeq z \ltsimeq 3.5$ (allowing foreground removal). Distributed over a core of diameter $D_{\rm core} = 300$ m 
(as the start of the SKA$_{2}$ mid-frequency AA core) this  would provide a core beam equivalent to $\sim 25$ comoving Mpc  (sufficient to Nyquist sample the first three BAO `wiggles'; see 
also Chang et al.\ \cite{Chang2008}) and $\mu_{\rm f} \sim 0.5$. This would provide an r.m.s. temperature sensitivity in a 1000-hr exposure of (assuming a
redshift depth of $\sim 25$ comoving Mpc, or $\approx 2.3$ MHz, and a system temperature $T_{\rm sys}$=50 K) $\approx 20 ~ \mu$K, meaning the thermal map noise will be $ \sim 3.5-$times lower than the
signals from HI structures, allowing mapping of these structures over, say, 20 deg$^{2}$ [$N_{\rm station} \gtsimeq 30$]. 
Over $3.5 \ltsimeq z \ltsimeq 6.0$, the data quality will be compromised by the sparse nature of the (low-frequency)
AA antennae at the relevant frequencies: detections via auto-correlation techniques might prove challenging, but are certainly not implausible.

These considerations show that, in going from SKA$_{1}$ to SKA$_{2}$, it will be critical to use the results of the AIP to optimally enhance the mapping speed of the $z \ltsimeq 2$ ($\gtsimeq 470$ MHz) Universe with the SKA. The gain achieved will be by a factor $\sim$100-10000, depending on the adopted AIP technology. Crudely,  if the eventual SKA$_{2}$ realization expands only the existing (70-450 MHz) SKA AA element (i.e. no mid-frequency AAs are built), a factor $\sim 10$ increase in the number of single-pixel-feed dishes would enhance the SKA mapping speed by `just' 100, meaning a 20,000 deg$^{2}$ 5$\sigma$-thresholded ($\sim 10^{9}$ objects to $z \ltsimeq 2$) survey would be unfeasible  
(Abdalla et al.\ \cite{Abdalla2010}): the best that could be managed would be a thresholded $\sim 100 ~ \rm deg^{2}$ survey to $z \sim 2$. This would be interesting as an adjunct, e.g. to the Euclid deep field, but would not constitute a game-changing experiment in cosmology, and would be uncompetitive with the 
wide-area near-IR redshift survey with Euclid. If the AIP selects an SKA$_{2}$ realisation
including $250 \times 56$-m diameter densely-packed AA stations\footnotemark (Rawlings \& Schilizzi \cite{Rawlings2011}),
\footnotetext{
The total antenna count in this mid-frequency aperture array would be, assuming $\lambda / 2$  sampling at 500 MHz ($\lambda=0.6 ~$m), 
$(250 \times \pi \times 56^{2}) / (4 \times (0.6 /2)^2) \sim 7 \times 10^{6}$, and $\sim (800/500)^2$ higher if, as preferred for imaging quality, the 500-800 MHz regime is entirely in the densely-packed regime
(Schilizzi et al.\ \cite{Schilizzi2007}). In broad-brush terms, these order-of-magnitude increases in antenna count (over SKA$_{1}$) look 
plausible within the SKA$_{2}$ funding envelope (Bolton et al.\ \cite{Bolton2009}). This is
certainly a case where crude scaling arguments have the potential to be dangerous, and the development of increasingly sophisticated costing tools up to the 2017 decision point on SKA$_{2}$ will become an increasingly crucial aspect of the project.
}
then the mapping speed gain would include a factor of $\sim 100$ from sensitivity increase, and a further factor of  $\sim 100$ from FOV increase (for mid-frequency AAs over dishes)\footnotemark , allowing `all sky' thresholded surveys (Table~\ref{tab:one}).  
\footnotetext{
The `half-way house' of using PAFs to expand the FOV of all the 15-m dishes is another possibility, but at 450 MHz ($\lambda = 0.67 ~ \rm m$), a phased array feed (with 10-100 beams) would become, in size, a large fraction of the primary reflector of the dish, and the cost of data transport and correlation may prove prohibitive (see Sec.\ \ref{sec:conc} for a potentially stronger cosmological science case for SKA$_{2}$ PAFs).
}

The resulting `billion-galaxy' surveys are needed to address questions such as neutrino mass, that is measurable to the lowest limit allowed by particle physics experiments at SKA$_{2}$ sensitivity (Abdalla \& Rawlings \cite{Abdalla2007}), and sub-per-cent accuracy on the dark energy $w$ parameter or demonstration, via measuring the cosmic evolution of the growth factor $g$, of the need for post-Einstein gravity. All such experiments would be well
serviced by SKA$_{2}$ galaxy power spectra (in several independent redshift bins, see Table~\ref{tab:one}) achieving high signal-to-noise ratio on BAOs and other features. They would also allow marginalization over galaxy bias through measurement of velocity-space distortions (Abdalla et al.\ \cite{Abdalla2010}). 

Additionally, with foreground removal using both low- and mid-frequency AAs, there are prospects of applying the intensity mapping method with both cores  to get a comparable measure of the $2 \ltsimeq z \ltsimeq 6$ power spectrum (Table~\ref{tab:one}) - this could be critical for distinguishing between effects of dark energy (that have little local effect  in the high-$z$ Universe) and other effects such as small, but non-zero, intrinsic Universal curvature. Statistical studies of the density field beyond $P(k)$, e.g. to high-order statistics or issues of primordial non-Gaussianity, can be applied to these large-number-of-mode surveys, and by testing competing models  of inflation (Jeong \& Komatsu 2009), will start to address the next set of key questions in cosmology.

\section{Concluding Remarks}
\label{sec:conc}

From Table~\ref{tab:one} we see that SKA is unique in future astronomy in allowing wide-field access to the very distant ($z \sim 6-30$) Universe when the 
first stars, galaxies and black holes formed and the Universe was reionized. The phased roll-out of the SKA means that its first results can be available alongside the first operation of the ELTs (and well-established facilities like ALMA, JWST etc) that will be available for studying objects in the EoR. However, the ways in which the HI fluctuations trace $P(k)$ (Equation~\ref{eqn:one}) at very high redshift are probably too complicated to add much to questions like the mass of neutrinos, the dark-energy $w$ parameter or post-Einstein gravity.

With regard to allowing substantially improved cosmological measurements over those already published, or those to be measured by 
BOSS and HETDEX, the most promising surveys appears to be those at $z < 6$ emphasised in bold text in Table~\ref{tab:one}
(see also Loeb \& Wyithe \cite{Loeb2008}).  Amongst planned experiments, only SKA, BigBOSS and Euclid have the ability to make cosmic-variance-limited measurements over the redshift range $0.2 \leq z \leq 2$ where there are $\sim 10^{7}$ modes available for study. The SKA surveys could have significantly higher effective volumes than Euclid or BigBOSS surveys (Kim et al.\ \cite{Kim2011}), and will probe much of this volume in the 
$n V P(k) \gg 1$ regime (Abdalla et al.\  \cite{Abdalla2010}). This means that analyses of SKA data will have the opportunity to
test the stability of results to how the tracers of the underlying density field are selected. It is of course highly plausible that the combination of SKA,
BigBOSS and Euclid datasets will prove much more powerful together than separately: the potential of eliminating systematics through cross-correlation may prove key to achieving the best-possible reductions in the error budgets on the cosmological parameters.

Here,  we have focussed on HI cosmology as it is HI and pulsar astronomy that drive the design of SKA$_{1}$. The SKA may, however, contribute to cosmology in other ways (e.g. Sutherland et al.\  \cite{Sutherland2011}):

{\it Weak gravitational lensing.} Although considered by Rawlings et al.\ (\cite{Rawlings2004}), this is generally not emphasised in the SKA science case
due to the strong competition from optical facilities, either ground-based (e.g. LSST) or space-based (Euclid). It remains true that SKA has the 
potential to combine the large-sky-area coverage of LSST, with the superb control of the point-spread function available only with space
missions at optical wavebands - and even then challenging if these observations are obtained with only limited colour information; SKA can also help with tomographic weak-lensing experiments using redshifts from its HI redshift surveys. 
SKA$_{1}$ will be a very useful weak lensing facility, and particularly so if the AIP yields a dish-FOV-extension technology opening up the possibility of $\gtsimeq 20 ~ \rm deg^{2}$ surveys at 1000-hr exposure depth at $\sim$1-2 GHz  frequencies where the PSF width is $\sim 0.5$-arcsec, as needed for weak-lensing experiments. In the spirit of the discussion of HI, it also remains plausible that the power of combining radio and optical experiments will be particularly crucial for pushing the determination of the cosmological parameters to the next level of accuracy. There are hints that this may be the case from the first attempts at joint radio-optical weak lensing that find stronger and cleaner signal in cross-correlation than in auto-correlation (Patel et al.\ \cite{Patel2010}).

{\it Measurement of $H_{0}$.} SKA studies of distant water masers for $H_{0}$ measurement was considered by Rawlings et al.\ (\cite{Rawlings2004}), with the caveat that the rest-frame frequency of water masers is $22$ GHz, so large ranges of redshift will be unavailable unless the SKA dishes are operated at frequencies between 10-22 GHz - this is currently a goal, rather than a requirement, of the SKA design process. Strong-gravitational-lensing experiments (e.g. Koopmans et al.\  \cite{Koopmans2004}) are also sensitive to $H_0$ through differential time delays (given a lens model encoding details of dark matter in collapsed structures): this provides an exciting combination of constraints on dark energy and dark matter 
that is worth a serious re-examination in the light of plans for SKA$_{1}$ (e.g.\ Heywood et al. \cite{Heywood2011b}).

\begin{acknowledgements}
The author thanks the EC for supporting the Crete meeting and the SKA project via the FP7 RadioNET and PrepSKA projects respectively. He
also thanks Arnold van Ardenne, 
Andy Faulkner, Ian Heywood, Gary Hill, Leon Koopmans, Francois Levrier and Mario Santos for discussions.
\end{acknowledgements}

\end{document}